\documentclass{autart}
\usepackage{amssymb}
\usepackage{graphicx} 
\usepackage{float}
\usepackage{setspace}
\usepackage{amsfonts}
\usepackage{amssymb,amsmath}
\usepackage{fancyhdr}
\usepackage{fancybox}
\usepackage{times}
\usepackage{color}
\usepackage[usenames,dvipsnames]{pstricks}
\usepackage{epsfig}
\usepackage{pst-grad} % For gradients
\usepackage{pst-plot} % For axes
\usepackage{blkarray}

\begin{document}

\begin{frontmatter}

\title{Global Synchronization of Pulse-Coupled Oscillators Interacting on Cycle Graphs\thanksref{footnoteinfo}}

 \thanks[footnoteinfo]{The material in this paper was not presented at any conference. This work was supported in part by CONICYT and the Fulbright Commission, and the Institute for Collaborative Biotechnologies under grants W911NF-09-D-0001 and W911NF-09-D-0001-0027. The content of the information does not necessarily reflect the position or the policy of the Government, and no official endorsement should be inferred.}
 \author[label1]{Felipe N\'u\~nez} \ead{fenunez@engineering.ucsb.edu},
 \author[label2]{Yongqiang Wang} \ead{wyqthu@gmail.com},
 %\author[label1]{Andrew R. Teel} \ead{teel@ece.ucsb.edu},
 \author[label1,label2]{Francis J. Doyle III}\ead{frank.doyle@icb.ucsb.edu}
 \address[label1]{Department of Electrical and Computer Engineering, University of California, Santa Barbara, California 93106-5080 USA}
 \address[label2]{Department of Chemical Engineering, University of California, Santa Barbara, California 93106-5080 USA}

\begin{abstract}
The importance of pulse-coupled oscillators (PCOs) in biology and engineering has motivated research to understand basic properties of PCO networks. Despite the large body of work addressing PCOs, a global synchronization result for networks that are more general than all-to-all connected is still unavailable. In this paper we address global synchronization of PCO networks described by cycle graphs. It is shown for the bidirectional cycle case that as the number of oscillators in the cycle grows, the coupling strength must be increased in order to guarantee synchronization for arbitrary initial conditions. For the unidirectional cycle case, the strongest coupling cannot ensure global synchronization yet a refractory period in the phase response curve is sufficient to enable global synchronization. Analytical findings are confirmed by numerical simulations.

\end{abstract}


\end{frontmatter}

% \linenumbers

%% main text
\section{Introduction}
\label{sec.int}
The synchronization of networks of agents has broad application in many fields, including: biological networks \cite{mir:90,sta:07}, mobile autonomous agents \cite{cor:06,sep:07}, and communication networks \cite{hon:05,pag:11}. In particular,  pulse-coupled oscillators (PCOs) are of great importance in biological and engineering systems since, despite its simple formulation, PCOs are able to model accurately complex network phenomena. Examples of biological systems modeled using PCOs include cardiac pacemakers \cite{pes:75}, and rhythmic flashing of fireflies \cite{buc:38}, while one of the most important applications of PCOs in engineering systems is time synchronization in sensor networks \cite{pag:11,hon:05,hu:06,kon:08,wan:11b,wan:12,wan:12b,nun:12}.

The synchronization of PCOs was first analyzed in the early work of Peskin \cite{pes:75}. In his work, Peskin made the following conjectures: 1) for arbitrary initial conditions, the system approaches a state in which all of the oscillators fire synchronously, and 2) this remains true, even when the oscillators are not identical. Numerous studies addressing these conjectures have been conducted, with variable success. In one of the most remarkable studies,  the authors proved that synchronization of identical PCOs in an all-to-all setting  is possible from every initial condition except from a set of zero Lebesgue measure \cite{mir:90}. Under the assumption of weak coupling, several authors have continued studying PCO networks using the phase model in \cite{izi:99} for different communication topologies and coupling functions. However, the weak coupling assumption needed to apply the techniques in \cite{izi:99}  makes it harder to prove a general result. Synchronization has proven difficult to establish and it is still not clear whether it is feasible or not to achieve global synchronization. Recently, \cite{mau:08,mau:11} showed that all-to-all connected PCO networks exhibit a dichotomic behavior, i.e., the network can either synchronize, or the oscillators form clusters distributed in the unit circle, depending on the characteristics of the phase response function. This dichotomic behavior is also present in networks with more general communication topologies as will be shown for the case of cyclic networks.  Interconnected oscillators interacting in cycles, or rings, have been used to model a variety of physiological phenomena such as segmental undulations in the leech, and hexapodal gait generation in insects \cite{dro:99}. Therefore, having a deep understanding of the synchronization properties of cyclic networks is of great interest for biologists and engineers. Along these lines, the authors in \cite{dro:99} presented insights on synchronization and phase-locking for unidirectional ring topologies by using a local linear approximation approach. However, a global result for cyclic networks, both uni- and bi-directional, of PCOs is not available.

The previous work on PCOs relied on the direct use of the biological model, which leads to a fixed feedback strategy. In this work we propose to re-design the PCO model to combine successful synchronization strategies taken from biology with modern control techniques to improve performance, as done by the authors in \cite{wan:11b,wan:12,wan:12b,nun:12,nun:12b,nun:12c}. Specifically, in this work, PCO networks are modeled as hybrid dynamical systems following our recent work \cite{nun:12,nun:12b,nun:12c} and the suggestion given in \cite{mau:12}. Networks of PCOs coupled through an optimal phase response curve (PRC) (in the sense of \cite{wan:12b}) and interacting on cyclic graphs are analyzed and necessary and sufficient conditions for global synchronization are given. As a side result, scenarios where a clustering behavior is observed are characterized, which highlights the existence of the dichotomic behavior previously mentioned. This dichotomy, sometimes undesirable, can be useful in particular systems such as wireless sensor networks. In fact, pulse-coupled time-division-multiple-access (TDMA) has been examined before for the all-to-all case with promising results \cite{deg:07}. The rest of this paper is organized as follows. Section \ref{sec.prelim} introduces preliminary concepts. In Section \ref{sec.2} a hybrid model for PCO networks and its main characteristics are presented. In section \ref{sec.3},  synchronization of PCO networks on cycle graphs is analyzed. Section \ref{sec.4} presents numerical experiments that confirm the analytical findings. Conclusions are given in Section \ref{sec.5}.
\section{Preliminaries}
\label{sec.prelim}
\subsection{Basic Notation and Definitions}
In this work, $\mathbb{R}$ denotes the real numbers, $\mathbb{R}_{\geq0}$ the set of nonnegative real numbers, $\mathbb{Z}_{\geq0}$ the set of nonnegative integers, $\mathbb{R}^n$ the Euclidean space of dimension $n$, and $\mathbb{R}^ {n\times n}$ the set of $n\times n$ square matrices with real coefficients. For a countable set $\chi$,  $|\chi|$ denotes its cardinality. For two sets $\Lambda_1$ and $\Lambda_2$, $\Lambda_1\setminus\Lambda_2$ denotes their difference. A set valued mapping $\varPhi: A \rightrightarrows B$ associates an element $\alpha\in A$ with a set $\varPhi(\alpha)\subseteq B$; the graph of $\varPhi$ is the set: $\text{graph}(\varPhi):=\{(\alpha,\beta)\in A\times B:\beta\in\varPhi(\alpha)\}$. $\varPhi$ is outer semi-continuous if and only if its graph is closed \cite{goe:09}.
\subsection{Hybrid Systems Preliminaries}
In this work we follow the hybrid systems framework given in \cite{goe:09}. A hybrid system $\tilde{\mathcal{H}}$ consists of continuous-time dynamics (flows), discrete-time dynamics (jumps), and sets on which these dynamics apply:
\begin{equation}
 \tilde{\mathcal{H}}:\begin{cases}
 \dot{x}\in F(x),&x\in\mathcal{C}\\
 x^+\in G(x),&x\in\mathcal{D}\\
         \end{cases}
\end{equation}
where the flow map $F$ and the jump map $G$ are set valued mappings, $\mathcal{C}$ is the flow set, and $\mathcal{D}$ is the jump set, $(F,\mathcal{C},G,\mathcal{D})$ is the data of the hybrid system $\tilde{\mathcal{H}}$. A subset $E\subset \mathbb{R}_{\geq 0}\times \mathbb{N}$ is a hybrid time domain if it is the union of infinitely many intervals of the form $[t_j,t_{j+1}]\times j$, or of finitely many such intervals. A solution to $\tilde{\mathcal{H}}$ is a function $\phi: \text{dom}\,\phi\to \mathbb{R}^n$ where $\text{dom}\,\phi$ is a hybrid time domain and for each fixed $j$, $t\mapsto \phi(t,j)$ is a locally absolutely continuous function on the interval $I_j=\{t:(t,j)\in \text{dom}\,\phi\}$. $\phi(t,j)$ is called a hybrid arc. A hybrid arc $\phi$ is nontrivial if its domain contains at least one point different from $(0,0)$, is maximal if it cannot be extended, and complete if its domain is unbounded.
\subsection{Graph formulation}
Consider a network of $N$ agents where $N\geq 4$. The communication between agents is modeled by a weighted directed graph $\mathcal{R} = \{\mathcal{V}, \mathcal{E_R} , \mathcal{A_R} \}$, where $\mathcal{V} = \{1,\ldots, N \}$ is the node set of the graph. $\mathcal{E_R} \subseteq \mathcal{V} \times \mathcal{V}$ is the edge set of the graph, whose elements are such that $(i,j) \in \mathcal{E_R}$ if and only if node $i$ can sense the state of node $j$. $\mathcal{A_R} = [a_{ij} ] \in \mathbb{R}^{N \times N}$ is the weighted adjacency matrix of $\mathcal{R}$ with $a_{ij} \geq 0$, where $a_{ij} > 0$ if and only if $(i,j) \in \mathcal{E_R}$.  In this work we focus on cycle graphs, w.l.o.g. we will consider the edge set given by $\mathcal{E_R}=(1,N)\cup\bigcup_{i=1}^{N-1}(i+1,i)$, i.e., node $i+1$ can sense the state of node $i$. Define $\bar{\mathcal{E}}_\mathcal{R}$ as the bidirectional, or undirected, version of $\mathcal{E_R}$ , i.e., if $(i,j)\in\mathcal{E_R}$ then $(i,j)$ and $(j,i)\in \bar{\mathcal{E}}_\mathcal{R}$. It should be noted that in the context of PCO networks $(i,j)\in\mathcal{E_R}$ means that node $i$ can sense the firing of node $j$ and thus $i$ updates its state after the firing of $j$.
\section{Pulse-coupled oscillator networks}
\label{sec.2}
\subsection{Model}
The network consists of $N$ oscillators interacting on the cycle graph $\mathcal{R} = \{\mathcal{V}, \mathcal{E_R} , \mathcal{A_R} \}$ or on its bidirectional version $\bar{\mathcal{R}}=\{\mathcal{V}, \bar{\mathcal{E}}_\mathcal{R} , \bar{\mathcal{A}}_\mathcal{R} \}$. Each oscillator modifies its phase following its natural frequency and using the information received in the form of pulses. Pulses are generated following an integrate-and-fire process, i.e., when its phase reaches the limit ($2\pi$ in this case), the oscillator emits a pulse and resets its phase to 0. When an oscillator receives a pulse, it updates its phase according to the coupling strength $l\in(0,1]$ and a function of its current phase value known as phase response curve, which is commonly used in the analysis of oscillatory biological systems and is formally defined in the framework of hybrid systems as follows:

\begin{defn}[Phase Response Curve]
 A phase response curve (PRC), or phase resetting curve \cite{can:10,ser:76}, describes the change in the phase of an oscillator resulting from a pulse stimulus as a function of the phase at which the pulse is received. A phase response curve $Q:[0,2\pi]\rightrightarrows \mathcal{Q}\subseteq\mathbb{R}_{\geq0}$ is called an advance-only PRC. A phase response curve $Q:[0,2\pi]\rightrightarrows \mathcal{Q}\subseteq \mathbb{R}$ such that there exists $q_1$, $q_2\in[0,2\pi]$  for which $\bar{q}_1\in Q(q_1),\;\bar{q}_1>0$ and $\bar{q}_2\in Q(q_2),\;\bar{q}_2<0$ is called an advance-delay PRC.
\end{defn}

In this work we consider a constant coupling strength $l$, and then the weighted adjacency matrices $\mathcal{A_R},\bar{\mathcal{A}}_\mathcal{R}$ are such that $a_{ij}\in\{0,l\}$. The network of $N$ oscillators is modeled by the hybrid system $\mathcal{H}$ with state $x$ defined as:
$$
 x:=[x_1,\ldots,x_N]^T\in[0,2\pi]^{N}
$$
where $x_i \in [0,2\pi]$ denotes the phase of the $i$th oscillator. 
The data of $\mathcal{H}$ is given by \cite{nun:12}:

If $x\in\mathcal{C}:=\{x\in\mathbb{R}^{N}: x_i\in[0,2\pi],\:\forall i\in\mathcal{V}\}$ $:=[0,2\pi]^{N}$ then:
\begin{equation}
 \dot{x}_i=w_i
\end{equation}
similarly, if $x\in\mathcal{D}_i:=\{x\in\mathcal{C}: x_i=2\pi\}$ then:
\begin{align}
 x_i^+&=0\\
 x_j^+&\in\begin{cases}
	    x_j+a_{ij}q, & \text{if}\; x_j+a_{ij}q\in (0,2\pi),\;q\in Q(x_j)\\
	    2\pi , & \text{if}\; x_j+a_{ij}q\geq2\pi,\;q\in Q(x_j)\\
	    0 , & \text{if}\; x_j+a_{ij}q\leq0,\;q\in Q(x_j)\\
         \end{cases}\nonumber
\label{eqn.model}
\end{align}
where $w_i\in \mathbb{R}_{>0}$ denotes the natural frequency, $a_{ij}\in\{0,l\}$ is the corresponding entry from $\mathcal{A_R}$ ($\bar{\mathcal{A}}_\mathcal{R}$), and $Q: [0,2\pi]\rightrightarrows \mathbb{R}$ is the phase response curve. We will assume identical natural frequencies, i.e., $w_i=w,\;\forall i\in\mathcal{V}$. It should be noted that the $\in$ in (3) implies that the PRC might be a set valued mapping. The jump map can be rewritten using the saturation function as:
\begin{align}
 x_i^+&=0\nonumber\\
 x_j^+&\in\text{sat}^{2\pi}_0(x_j+a_{ij}Q(x_j)),\;\;x\in \mathcal{D}_i
\end{align}
where $\text{sat}^{2\pi}_0$ is the linear function with slope one that saturates at $2\pi$ from above and $0$ from below. Moreover, the effect of the saturation function can be eliminated by imposing a range condition on the PRC as $\text{graph}(Q)\subseteq\Omega:=\{(x,y):x\in[0,2\pi],-x\leq y\leq2\pi-x\}$. This condition is not restrictive since if part of the graph lies outside $\Omega$, we can replace the PRC with a saturated version of it, without affecting the resulting dynamics. To continue the analysis we utilize the following assumption.
\begin{assum}
\label{ass.3}
 The PRC $Q$ is such that: $Q(0)=Q(2\pi)=0$. Moreover, $Q$ is an outer semi-continuous set-valued mapping and locally bounded on $\mathcal{D}_i,\;\forall i\in\mathcal{V}$.
\end{assum}
Finally, the jump set is defined as the union over the node set of the individual jump sets previously defined:  
\begin{equation}
 \mathcal{D}:=\bigcup_{i\in\mathcal{V}} \mathcal{D}_i
\end{equation}
It should be noted that the proposed model is able to handle multiple oscillators firing at the same time. Assumption \ref{ass.3} guarantees that the hybrid system $\mathcal{H}$ with data $(\mathcal{C},F,\mathcal{D},G)$ as defined above is well-posed \cite{goe:09}.

\begin{rem}
 An important concept used in the analysis of PCOs is absorption \cite{mau:08,mir:90}, which leads to synchronization in finite time. It should be noted that in our model (2)-(4), absorption only takes place when $l=1$; hence, synchronization will occur in finite time only when $l=1$ and otherwise it will be asymptotic.
\end{rem}

\subsection{Solutions to the hybrid model}
The behavior of the solutions to the hybrid system $\mathcal{H}$ on a cycle graph, either $\bar{\mathcal{R}}$ or $\mathcal{R}$, is characterized as follows. 

\begin{prop}
\label{prop.1}
 For every initial condition $\phi_0\in\mathcal{C}\cup\mathcal{D}$, there exists a nontrivial solution starting at $\phi_0$. Furthermore, let $\phi$ be a maximal solution to the hybrid system $\mathcal{H}$ on $\bar{\mathcal{R}}$ ($\mathcal{R}$) with initial condition $\phi(0,0)=\phi_0\in \mathcal{C}\cup \mathcal{D}$. Then the following statements are true:
 \begin{enumerate}
  \item[(a)] $\phi$ is complete.
   \item[(b)] $\phi$ has at most $N$ consecutive jumps with no flow in between. 
  \item[(c)] The amount of ordinary time between jumps is at most $\tfrac{2\pi}{w}$.
\end{enumerate}
\end{prop}
The proof is given in the Appendix.
\begin{rem}
 Proposition \ref{prop.1} tells us that solutions behave as observed in biological systems and fulfill reasonable engineering expectations: they are complete and jump periodically. Statement (b) rules out the existence of Zeno solutions and statement (c) guarantees that jumps are persistent, i.e., it rules out the existence of solutions that only flow. It should also be noted that, in general, solutions to $\mathcal{H}$ are not unique. %Moreover, statement (d) ensures that the period of every oscillator is upper and lower bounded.
\end{rem}
\section{Global Synchronization of PCO networks on Cycle Graphs}
\label{sec.3}
In this section we analyze the synchronization properties of PCO networks interacting on a cycle graph.  Synchronization is characterized as convergence to a compact set. Note that even though ``easy'' initial conditions can synchronize under weaker conditions, the following results give the weakest conditions for global synchronization.

Define the synchronization set as $\mathcal{S}:=\{x\in\mathcal{C}:|x_i-x_{i+1}|=0,\text{ or }|x_i-x_{i+1}|=2\pi,\, \forall i\in\mathcal{V}\}$, with the understanding that node $N+1$ is mapped to node $1$ (and node $0$ to node $N$ in the following analysis). We will say that the network synchronizes if the state $x$ converges to the set $\mathcal{S}$. %In the remainder of this section we will study the stability properties of this set.

Consider the following family of functions representing the distance to the synchronization set $\mathcal{S}$:
\begin{equation}
 v_{i,i+1}(x)=\min\left(|x_i-x_{i+1}|,2\pi-|x_i-x_{i+1}|\right)
\end{equation}
note that $v_{i,i+1}(x)$ represents the length of the shortest segment joining oscillators $i$ and $i+1$. Define the vector of distance functions as:
\begin{equation}
 V:=[v_{1,2},v_{2,3},\ldots,v_{N-1,N},v_{N,1}]^T\in[0,\pi]^N
 \label{eqn.v}
\end{equation}
and the length of the cycle as $\mathbf{1}^TV$, where $\mathbf{1}$ is the $N$-dimensional column vector of all ones. We will refer to the component $i\in\{1,2,\ldots,N\}$ of $V$ as $V_i$. These components are continuous functions with respect to $x$, and positive definite with respect to $\mathcal{S}$. It is clear, since the oscillators have identical natural frequencies, that $V_{i}$ remains unchanged during flows. Hence, the discrete-time dynamics (jumps) entirely determine the synchronization properties of the system. We will analyze the convergence properties of the underlying discrete-time system to prove synchronization, i.e., we focus on the difference inclusion (4).

In the following, we consider that the feedback is given by the optimal advance-delay PRC:
\begin{equation}
 Q(x)=\begin{cases}
	    2\pi-x, & \text{if}\; x>\pi\\
	    \{\pi,-\pi\}, & \text{if}\; x=\pi\\
	    -x, & \text{if}\; x<\pi\\
         \end{cases}
\label{eqn.prc2}
\end{equation}
which corresponds to the set-valued regularization of the discontinuous function $Q(x)=2\pi-x, \,x\in[\pi,2\pi];\;Q(x)=-x, \,x\in[0,\pi)$. Note that (\ref{eqn.prc2}) is an outer semi-continuous set-valued mapping and locally bounded; hence Assumption \ref{ass.3} holds. Moreover, the graph of (\ref{eqn.prc2}) lies entirely inside the set $\Omega$. The PRC (\ref{eqn.prc2}) has been proven to be optimal in terms of synchronization rate in our earlier work \cite{wan:12b} and thus, it will be used in this work. 

Before stating the synchronization results, we need to introduce the concept of refractory period and a technical lemma that can be easily derived from Theorem 1 in \cite{wan:12}. 

\begin{defn}[Refractory period]
 A refractory period is an interval $[0,r]\subseteq[0,2\pi]$, where $r$ is the length of the refractory period, such that if the phase of an oscillator is inside the interval, it does not react to an incoming pulse, i.e., a refractory period of length $r$ corresponds to a dead zone in the PRC in the interval $[0,r]$ \cite{ser:76}.
\end{defn}

\begin{lem}
\label{lem.synch}
 Consider a network of PCOs interacting on a cycle graph, either $\mathcal{R}$ or $\bar{\mathcal{R}}$ . If the initial phases are such that $$\max_{i,k\in\mathcal{V}}|x_i(0,0)-x_k(0,0)|<\pi,$$ $l\in(0,1]$, and the PCR is given by (\ref{eqn.prc2}), then the network converges asymptotically to the set $\mathcal{S}$ even if there exists a refractory period in the PRC of length $r\leq\pi$. 
\end{lem}

The following Theorems are the main results of this paper and provide necessary and sufficient conditions for global synchronization of PCOs interacting on cycle graphs.

 \begin{thm}
\label{thm.1}
  Consider the network of PCOs with dynamics $\mathcal{H}$ interacting on $\bar{\mathcal{R}}$, and with PRC given by (\ref{eqn.prc2}). The network synchronizes from every initial condition if and only if the coupling strength $l$ is larger than the critical coupling $l^*$, which is given by:
  \begin{equation}
   l^*=\frac{N}{2}-\frac{\sqrt{N^2-4(N-2)}}{2}
  \end{equation}
 \end{thm}

A similar condition can be derived for the unidirectional graph $\mathcal{R}$. 

 \begin{thm}
\label{thm.2}
  Consider the network of PCOs with dynamics $\mathcal{H}$ interacting on $\mathcal{R}$, and with PRC given by (\ref{eqn.prc2}). Moreover, consider that there exists a refractory period of length $r=\pi$ in the PRC of 1 oscillator. The network synchronizes from every initial condition if and only if the coupling strength $l$ is larger than the critical coupling $l^*$, which is given by:
  \begin{equation}
   l^*=\frac{N-2}{N-1}
  \end{equation} 
 \end{thm}

To prove Theorems \ref{thm.1} and \ref{thm.2} we rely on the following Lemma.

\begin{lem}
\label{lem.v}
 Consider the distance vector $V$ defined in (\ref{eqn.v}) and the length of the cycle defined as $\mathbf{1}^TV$. At any time instant $(\bar{t},\bar{j})$, let $i^*\in\mathcal{V}$ be the index of the oscillator with the largest phase and $i_*\in\mathcal{V}$ the index of the oscillator with the smallest phase. Define $\mathcal{U}_1:=\{x\in\mathcal{C}: x_i\geq x_{i+1}\forall i\in\mathcal{V}\setminus \{i_*\} \}\cap \{x\in\mathcal{C}: \mathbf{1}^TV=2\pi\}$, $\mathcal{U}_2:=\{x\in\mathcal{C}: x_i\leq x_{i+1}\forall i\in\mathcal{V}\setminus \{i^*\} \}\cap \{x\in\mathcal{C}: \mathbf{1}^TV=2\pi\}$, and $\mathcal{U}:=\mathcal{U}_1\cup\, \mathcal{U}_2$. The following claims hold:%and define $\bar{i}:=\min\{i^*,i_*\}$
 \begin{enumerate}
  \item[(a)] If $\mathbf{1}^TV<2\pi$, then $\exists$ $i\in\mathcal{V}\setminus \{i^*,i_*\}$ such that $|x_{i}-x_{i+1}|>\pi$, or $|x_{i^*}-x_{i_*}|<\pi$
  \item[(b)] If $\mathbf{1}^TV>2\pi$, then $\exists$ $i\in\mathcal{V}$ such that when $x_i=2\pi$ we have that $x_{i+2}\in[0,x_{i+1})$ and $x_{i+1}\leq\pi$, or $x_{i+2}\in(x_{i+1},2\pi]$ and $x_{i+1}\geq\pi$, or $|x_{i+2}-x_{i+1}|>\pi$; hence $\mathbf{1}^TV$ decreases after $i$ jumps.
  \item[(c)] If $\mathbf{1}^TV=2\pi$ and $x\notin\mathcal{U}$, then there exists $i\in\mathcal{V}$ such that $\mathbf{1}^TV$ decreases after $i$ jumps.
  \item[(d)] If $x\in\mathcal{U}$, then $|x_{i^*}-x_{i_*}|\geq\pi$ and $|x_{i}-x_{i+1}|\leq\pi,\,\forall i\in\mathcal{V}\setminus \{i^*,i_*\}$.
 \end{enumerate}
\end{lem}

The proof of Lemma \ref{lem.v} is given in the appendix.

\begin{rem}
  Note that Lemma \ref{lem.v}(a) implies that if $\mathbf{1}^TV<2\pi$, then conditions of Lemma \ref{lem.synch} hold up to a rigid rotation of the oscillators. Hence, when $\mathbf{1}^TV<2\pi$ the network always synchronizes. Moreover,  conditions in Lemma \ref{lem.synch} and Lemma \ref{lem.v}(a) imply that the oscillators are in a semicircle, a problem equivalent to a consensus problem in $\mathbb{R}^N$ \cite{mau:12b}.
 \end{rem}

  \begin{rem}
  Statement $(b)$ of Lemma \ref{lem.v} means that when $\mathbf{1}^TV>2\pi$, the length will eventually decrease. Regarding global synchronization, initial conditions for which $\mathbf{1}^TV>2\pi$ do not represent a problem since in these cases the length will decrease. In fact, we will show that the only problematic situation is when $x(0,0)\in\mathcal{U}$. 
 \end{rem}
 
 Now we proceed to prove Theorems \ref{thm.1} and \ref{thm.2}.
 
 \begin{pf}[Theorem \ref{thm.1}]
 To prove sufficiency, the strategy is to show that every solution is such that eventually $\mathbf{1}^TV<2\pi$ and hence Lemma \ref{lem.synch} yields synchronization of the network. Consider an arbitrary initial condition $x(0,0)\in\mathcal{C}$, we have four possible scenarios:

\emph{i) $x(0,0)\in\mathcal{C}: \mathbf{1}^TV<2\pi$}

In this case, directly applying Lemma \ref{lem.synch} guarantees synchronization.

\emph{ii) $x(0,0)\in\mathcal{C}: \mathbf{1}^TV>2\pi$}

Lemma \ref{lem.v}(b) guarantees that $\mathbf{1}^TV$ will decrease while $\mathbf{1}^TV>2\pi$, then there exists a time instant $(t_{ii},j_{ii})$ such that either $\mathbf{1}^TV=2\pi,x(t_{ii},j_{ii})\notin\mathcal{U}$ or $x(t_{ii},j_{ii})\in\mathcal{U}$. At this point we can reinitialize the system in case \emph{iii)} or \emph{iv)}.

\emph{iii) $x(0,0)\in\mathcal{C}: \mathbf{1}^TV=2\pi,x(0,0)\notin\mathcal{U}$}

Lemma \ref{lem.v}(c) ensures that the length will decrease and then there exists a time instant $(t_{iii},j_{iii})$ at which $\mathbf{1}^TV<2\pi$. At this point we can reinitialize the system in case \emph{i)} and invoking Lemma \ref{lem.synch} gives synchronization.

\emph{iv) $x(0,0)\in\mathcal{U}$}
 
In this case, the situation is more complicated. To show that the system jumps outside $\mathcal{U}$, we analyze the change in $V$ when an oscillator $i$ jumps.
Consider $x\in\mathcal{D}$, which is the union of the jump conditions for all $x_i$, suppose w.l.o.g.that node $i$ is about to fire, denote the time as $(t,j)$ and the state as $x(t,j)$. We drop the time indices $t$ and $j$ to facilitate the notation; however, the reader should be aware that the time domain is a hybrid one, that $V^+(x)$ means $V(x(t,j+1))$, and $V_0=V(x(0,0))$. We have that $x_i=2\pi$ and $x_{i+1}\in[0,2\pi]$, then:
\begin{align}
 V_{i}(x)&=\min\left\lbrace 2\pi-x_{i+1},x_{i+1}\right\rbrace\\
 V^+_{i}(x)&= x_{i+1}(1-l)\text{ or } (2\pi-x_{i+1})(1-l)\nonumber
\end{align}
depending on whether $x_{i+1}\in[0,\pi]$ or $x_{i+1}\in[\pi,2\pi]$. Then, $V_{i}(x)>V^+_{i}(x)=(1-l)V_{i}(x)$ holds for any value of $x_{i+1}$ before $x_i$ jumps. Note that since the previous analysis is valid for all $i$ we have  $V_{i-1}(x)>V^+_{i-1}(x)=(1-l)V_{i-1}(x)$. Next we analyze the change in $V_{i+1}$. In this case we have:
\begin{align}
 V_{i+1}(x)&=\min\left\lbrace|x_{i+1}-x_{i+2}|,2\pi-|x_{i+1}-x_{i+2}|\right\rbrace\nonumber\\
 V^+_{i+1}(x)&=\min\left\lbrace|x_{i+1}-x_{i+2}+lQ(x_{i+1})|,\right.\\
 & \left. 2\pi-|x_{i+1}-x_{i+2}+lQ(x_{i+1})|\right\rbrace\nonumber
\end{align}
 Since $x(t,j)\in\mathcal{U}$, the phase ordering (either $x_i\geq x_{i+1}$ or $x_i\leq x_{i+1}$)  and $|x_{i}-x_{i+1}|\leq\pi$ from Lemma \ref{lem.v}(d) ensure that $V^+_{i+1}(x)= V_{i+1}(x)+lV_{i}(x)$ and $V^+_{i-2}(x)= V_{i-2}(x)+lV_{i-1}(x)$ hold, provided $V_{i+1}(x)+lV_{i}(x)<\pi$ and $V_{i-2}(x)+lV_{i-1}(x)<\pi$ (note that if the previous conditions do not hold, the length decreases and since $\mathbf{1}^TV<2\pi$, the network synchronizes). The other components of $V$ remain unchanged when $i$ jumps. We can then write the change of $V$ after $i$ jumps in matrix form by using the following transition matrices
 \begin{tiny}
\begin{equation}
\bar{C}_i=\bordermatrix{
& & & & &i^{th}\cr
&1 &  0  & \cdots & 0&0&0&0&\cdots&0\cr
& 0  & 1 & \cdots & \vdots&\vdots&\vdots&\vdots&\cdots&0\cr
& 0 & 0 & \ddots & l&\vdots&\vdots&\vdots&\cdots&0\cr
& 0 & 0 & \cdots & (1-l)&0&\cdots&\vdots&\cdots&0\cr
i^{th}&\vdots& \vdots & \cdots & 0 &(1-l)&0&\vdots&\cdots&0\cr
& \vdots & \vdots &\cdots& \vdots &l&1&\vdots&\cdots&0\cr
& \vdots & \vdots & \cdots &0&\vdots&0&\ddots&\cdots&0\cr
& \vdots & \vdots & \cdots &\vdots&\vdots&\vdots&\vdots&\ddots&\vdots\cr
& 0 & 0 & \cdots &0&0&0&0&\cdots&1}
\nonumber
\end{equation}
\end{tiny}
 Then, when $V_{i+1}(x)+lV_{i}(x)<\pi$ and $V_{i-2}(x)+lV_{i-1}(x)<\pi$ holds, the value of $V$ after $i$ jumps is given by $V^+=\bar{C}_iV$.   Note that  $\bar{C}_i$ are column stochastic matrices and then when $V^+=\bar{C}_iV$, $\mathbf{1}^TV^+=\mathbf{1}^TV$ holds,  i.e., the length remains constant and the state remains in $\mathcal{U}$. In the following, we will use an auxiliary system $\tilde{V}^+=\bar{C}_i\tilde{V}$ with $\tilde{V}_i\in\mathbb{R}$ and $\tilde{V}_0=V_0$ (note that the elements of $\tilde{V}$ are not restricted to $[0,\pi]$ as the elements of $V$)  to show that if $l>l^*$, the state will jump out of $\mathcal{U}$ and the network will synchronize.  
 It is a well known fact from consensus theory \cite{cao:08} that an infinite product of column stochastic matrices with positive diagonal entries, as $\bar{C}_i$, converges exponentially to a matrix of the form $\gamma \mathbf{1}^T$, where $\gamma$ is a column vector such that $\mathbf{1}^T\gamma=1$ \cite{cao:08}. By exploiting the particular structure of the $\bar{C}_i$ matrices, we can determine exactly the value of the vector $\gamma$ as follows. Assume the system $\tilde{V}^+=\bar{C}_i\tilde{V}$ is at the equilibrium $\tilde{V}^*=\gamma \mathbf{1}^T\tilde{V}_0=2\pi\gamma$ and w.l.o.g. $x\in\mathcal{U}_1$ and oscillator 1 is about to fire (note that in the bidirectional case, $x\in\mathcal{U}_1$ and $x\in\mathcal{U}_2$ are equivalent in terms of $\tilde{V}$). The phase ordering ensures that the firing sequence will be $1,2,\ldots,N$ and since the system is at equilibrium, the $C_i$ matrices induce a hard rotation on the elements of $\gamma$ (since the length cannot decrease). Hence, assuming $l\in(0,1)$,  the vector $\gamma$ must contain $N-2$ identical elements $\delta$, one element equal to $(1-l)\delta$ and one element equal to $\frac{\delta}{(1-l)}$. Moreover, we have that
\begin{equation}
 (N-2)\delta+(1-l)\delta+\frac{\delta}{(1-l)}=1
\end{equation}
holds. Since $l>l^*=\frac{N}{2}-\frac{\sqrt{N^2-4(N-2)}}{2}$, solving for $\frac{\delta}{(1-l)}$ gives $\frac{\delta}{(1-l)}>\frac{1}{2}$. Hence, if $l>l^*$ we have that, at the equilibrium, $\max \tilde{V}_i=\frac{\delta}{(1-l)}2\pi>\pi$. Then, a component of $\tilde{V}$ will converge exponentially fast \cite{cao:08} to a value larger than $\pi$, which in the original system, where $V_{i}\in[0,\pi]$, has to be interpreted as $|x_{I}-x_{I+1}|>\pi$ for some $I\in\mathcal{V}$ and hence we have $\mathbf{1}^TV<2\pi$. At this point, we can take this as initial condition for Lemma \ref{lem.synch}. It should be noted that if $l=1$, $\gamma$ contains only one non zero entry $\delta=1$, which ensures synchronization.  Hence, the network synchronizes from every initial condition $x(0,0)\in\mathcal{C}$.
The `only if' part follows easily by contradiction. First suppose that the network synchronizes from every initial condition and that $l\leq l^*$. Define the set $\bar{\mathcal{U}}^*:=\left\lbrace x\in\mathcal{U}:V_{i^*-1}=\frac{\delta}{(1-l)},V_{i_*}=(1-l)\delta,V_i=\delta\right.$ $\left.\forall i\in\mathcal{V}\setminus\{i^*-1,i_*\}\right\rbrace$ as the ``worst case'' set (note that this set contains the equilibrium of the system $\tilde{V}^+=\bar{C}_i\tilde{V}$) the result follows by using $x(0,0)\in\bar{\mathcal{U}}^*$ as a counterexample.\qed
\end{pf}

\begin{pf}[Theorem \ref{thm.2}]
The proof uses the same arguments as the proof of Theorem \ref{thm.1}. Cases \emph{i)}, \emph{ii)}, and \emph{iii)} follows the same arguments, yet case \emph{iv)} is different. To show that the system jumps outside $\mathcal{U}$,  first consider $x(0,0)\in\mathcal{U}_1$. In this case we have that when $i$ fires, the phase ordering and Lemma \ref{lem.v}(d) ensure that $x_{i+1}\in[\pi,2\pi]$ and then the refractory period has no effect. Following the same reasoning as the one for the proof of Theorem \ref{thm.1}, the transition matrices are given by  
\begin{tiny}
\begin{equation}
C_i=\bordermatrix{
& & & & &i^{th}\cr
&1 &  0  & \cdots & 0&0&0&0&\cdots&0\cr
& 0  & 1 & \cdots & \vdots&\vdots&\vdots&\vdots&\cdots&0\cr
& 0 & 0 & \ddots & 0&\vdots&\vdots&\vdots&\cdots&0\cr
& 0 & 0 & \cdots & 1&0&\cdots&\vdots&\cdots&0\cr
i^{th}&\vdots& \vdots & \cdots & 0 &(1-l)&0&\vdots&\cdots&0\cr
& \vdots & \vdots &\cdots& \vdots &l&1&\vdots&\cdots&0\cr
& \vdots & \vdots & \cdots &0&\vdots&0&\ddots&\cdots&0\cr
& \vdots & \vdots & \cdots &\vdots&\vdots&\vdots&\vdots&\ddots&\vdots\cr
& 0 & 0 & \cdots &0&0&0&0&\cdots&1}
\nonumber
\end{equation}
\end{tiny}
Note that the transition matrices are different from the bidirectional case due to the unidirectional nature of the graph. However, $C_i$ are also column stochastic matrices and hence their infinite product converges exponentially to a matrix of the form $\gamma\mathbf{1}^T$. We will again consider an auxiliary system $\tilde{V}=C_i\tilde{V}$ to prove that the system jumps outside $\mathcal{U}_1$. Since $x\in\mathcal{U}_1$, the phase ordering ensures that the firing sequence will be $1,2,\ldots,N$ and for the matrices $C_i$ to induce a hard rotation on $\tilde{V}$ at the equilibrium, assuming $l\in(0,1)$, the vector $\gamma$ must contain $N-1$ identical elements $\delta$ and one element equal to $\frac{\delta}{(1-l)}$. Moreover, we have that
\begin{equation}
 (N-1)\delta+\frac{\delta}{(1-l)}=1
\end{equation}
holds. Since $l>l^*=\frac{N-2}{N-1}$, solving for $\frac{\delta}{(1-l)}$ gives $\frac{\delta}{(1-l)}>\frac{1}{2}$. Hence, if $l>l^*$ we have that, at the equilibrium, $\max \tilde{V}_i=\frac{\delta}{(1-l)}2\pi>\pi$, Then, a component of $\tilde{V}$ will converge exponentially fast to a value larger than $\pi$, which in the original system, where $V_{i}\in[0,\pi]$, has to be interpreted as $|x_{I}-x_{I+1}|>\pi$ for some $I\in\mathcal{V}$ and hence we have $\mathbf{1}^TV<2\pi$. At this point, we can take this as initial condition for Lemma \ref{lem.synch}. It should be noted that if $l=1$, $\gamma$ contains only one non zero entry $\delta=1$, which ensures synchronization. Hence, the network synchronizes from every initial condition $x(0,0)\in\mathcal{U}_1$.

Now consider $x(0,0)\in\mathcal{U}_2$, and w.l.o.g. that $N$ will fire first. Suppose further that there is no refractory period. Note that in this case, the phase ordering ensures that the firing sequence will be $N,N-1,\ldots,1$. Hence, to ensure a hard rotation at the equilibrium, the vector $\gamma$ must contain $N-1$ identical elements $\delta$ and one element equal to $(1-l)\delta$. Moreover, we have that
\begin{equation}
 (N-1)\delta+(1-l)\delta=1
\end{equation}
holds. Then, the maximum feasible value for $\delta$ is $\frac{1}{N-l}$ and the network cannot synchronize, even if $l=1$. However, when there is a refractory period in one node the network can synchronize. Recall that, due to the phase ordering, nodes get pulses when their phases are in $[0,\pi)$. Then, if the refractory period is in node $i$, when node $i-1$ jumps $V_{i-1}$ is not affected; yet when node $i-2$ jumps, node $i-1$ is affected and $V_{i-1}$ is increased by $lV_{i-2}$. Therefore after one round of firings $V_{i-1}$ will have been increased by $lV_{i-2}$. Iterating this argument, $|x_{i-1}-x_{i}|>\pi$ will hold after a finite number of firing rounds, node $i$ will react to node's $i-1$ firing event, and $x_i\in[0,\pi)\;\forall i\in\mathcal{V}$. Invoking again Lemma \ref{lem.synch} completes the proof. 

The `only if' part follows by contradiction supposing that the network synchronizes from every initial condition and that $l\leq l^*$. Using the ``worst case'' initial condition $x(0,0)\in\mathcal{U}_1^*$ as counterexample, where $\mathcal{U}_1^*:=\left\lbrace x\in\mathcal{U}_1:V_{i^*}=\frac{\delta}{(1-l)},V_i=\delta\;\forall i\in\mathcal{V}\setminus\{i^*\}\right\rbrace$, yields a contradiction. \qed
\end{pf}
\begin{rem}
 The beneficial effects of a refractory period on the stability of PCO networks have been mentioned before \cite{hon:05,wan:12}. In the same sense, Theorem \ref{thm.2} states that the introduction of a refractory period enables global synchronization in the unidirectional case. It should be noted, however, that if more than one oscillator is affected by a refractory period, global synchronization cannot be guaranteed. To see this fact, suppose we have a $N$-node bidirectional cycle, where 2 nodes have a refractory period of length $r=\pi$. Consider an initial condition given by 2 clusters, one at $\pi$ and the other at $2\pi$ and suppose $l=1$. It can be derived that the oscillators containing the refractory period will remain $\pi$ apart while the other oscillators will jump back and forth.
\end{rem}

\section{Numerical Experiments}
\label{sec.4}
\begin{figure}
\begin{center}\includegraphics[
width=0.38\textwidth,keepaspectratio]{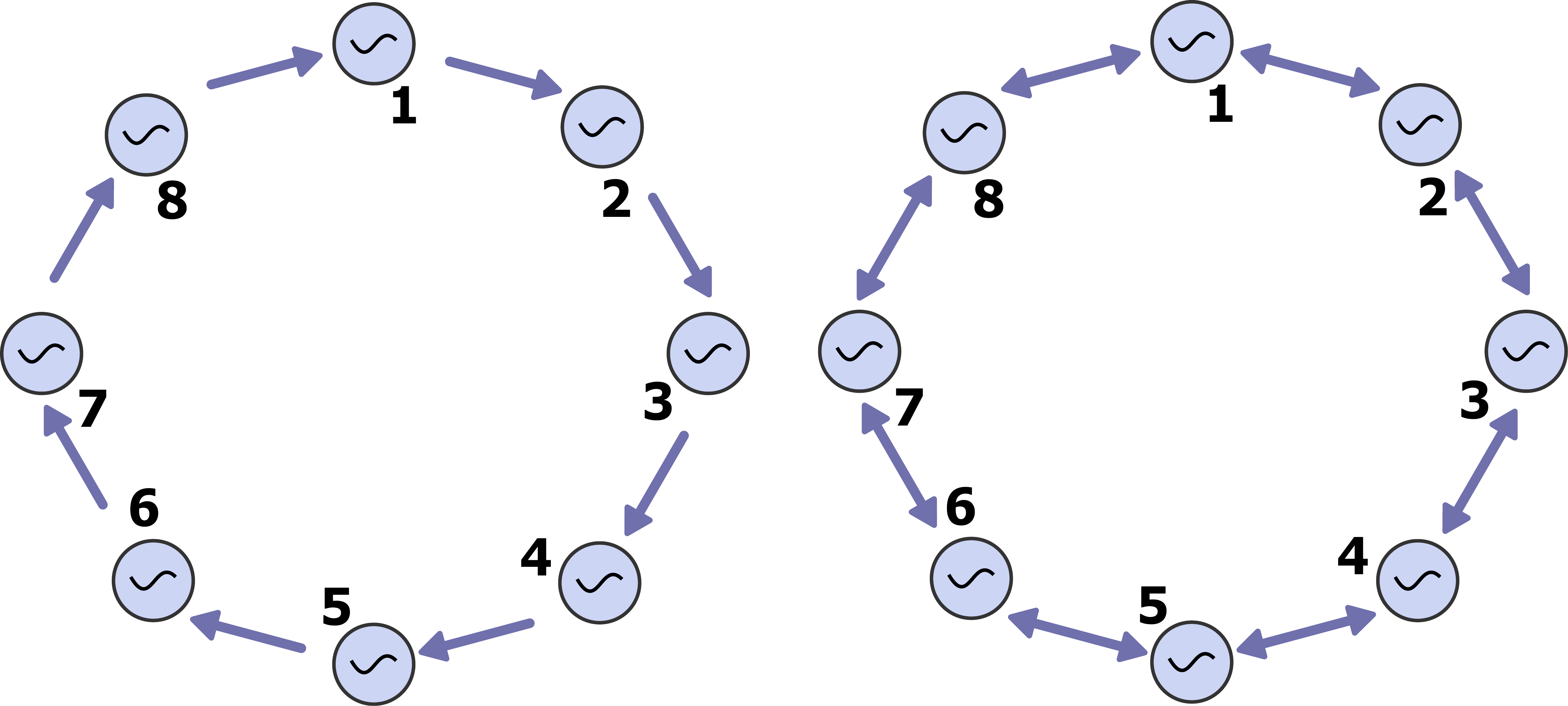}\end{center}
\caption{Network topologies used in the numerical experiments. Left: unidirectional ring of 8 nodes. Right: the bidirectional, or undirected, version of the ring of 8 nodes. Natural frequencies were set to $w=2\pi$ for all the experiments.}
\label{net.dist}
\end{figure}

To illustrate our analytical findings, several numerical experiments were conducted using a general hybrid systems simulator \cite{san:08}. Figure \ref{net.dist} shows the PCO networks used in the simulations consisting of 8 oscillators interacting on a bidirectional and a unidirectional graph. For all the experiments, natural frequencies were set to $w_i=w=2\pi$. 

\begin{figure}
\begin{center}\includegraphics[
width=0.48\textwidth,keepaspectratio]{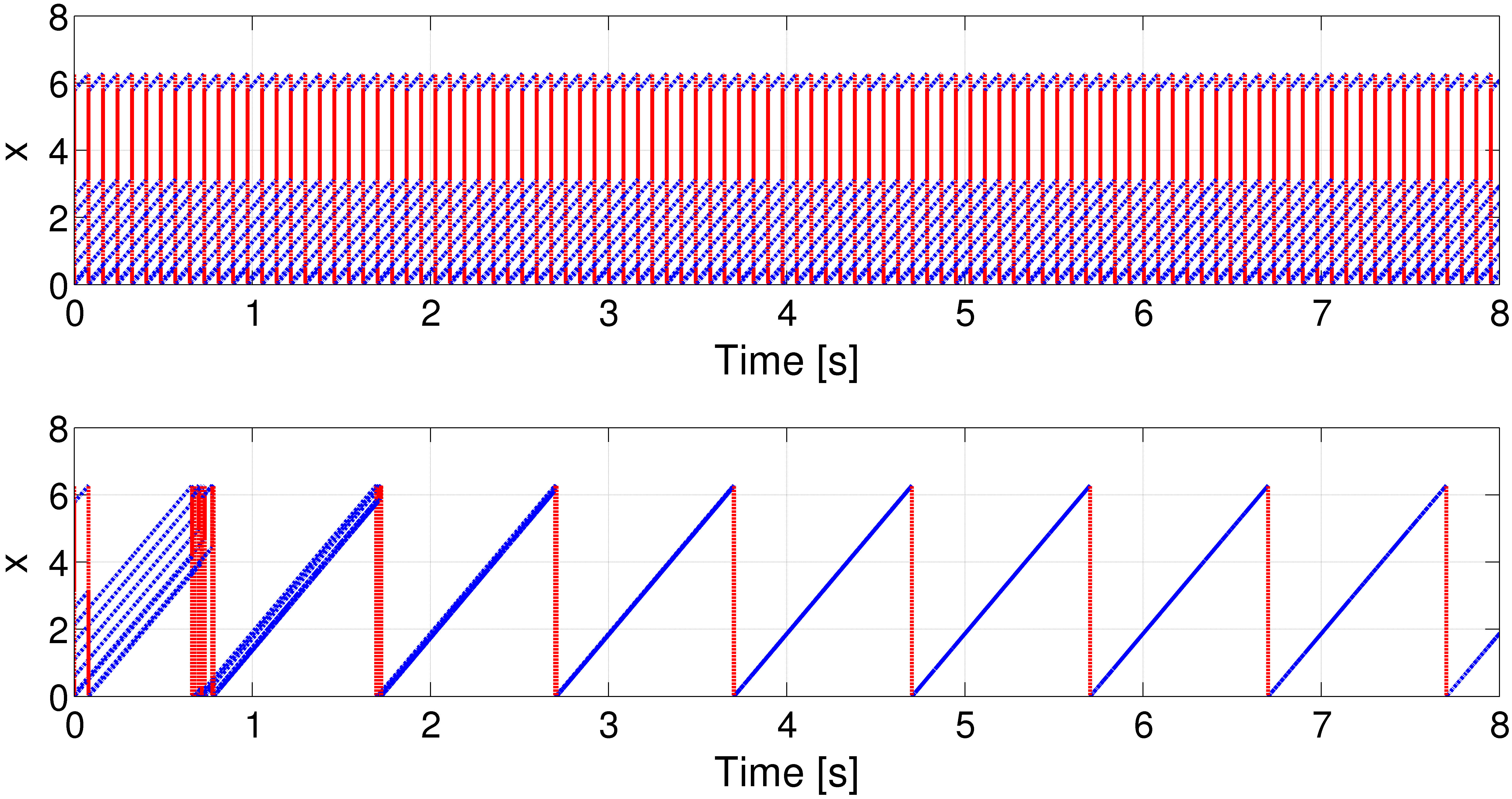}\end{center}
\caption{Simulation results for the \textbf{bidirectional} ring of figure \ref{net.dist} and initial condition $x(0,0)\in\bar{\mathcal{U}}^*$. Red lines denote jump instants and blue lines denote phase values. On the top plot $l=0.8377$; since from Theorem \ref{thm.1} we have $l^*=0.83772$ the network cannot synchronize. On the bottom plot $l=0.8378$; since in this case $l>l^*$ the network synchronizes.}% from the worst case, and thus globally from any initial condition.}
\label{res.bidir}
\end{figure}

Figure \ref{res.bidir} shows the results for the bidirectional graph with initial condition $x(0,0)\in\bar{\mathcal{U}}^*$. Solving the condition in Theorem \ref{thm.1} gives a critical coupling strength of $l^*=0.83772$. In the top plot the coupling strength is set below the critical value as $l=0.8377$; hence, the network cannot synchronize and the oscillators distribute in the interval $[0,2\pi]$. It can also be seen in the figure, that $\bar{\mathcal{U}}^*$ is in fact a TDMA-like equilibrium for the system.  On the other hand, when the coupling strength is increased to $l=0.8378$, i.e., above the critical value, the network asymptotically synchronizes, as shown in the bottom plot of Figure \ref{res.bidir}

\begin{figure}
\begin{center}\includegraphics[
width=0.48\textwidth,keepaspectratio]{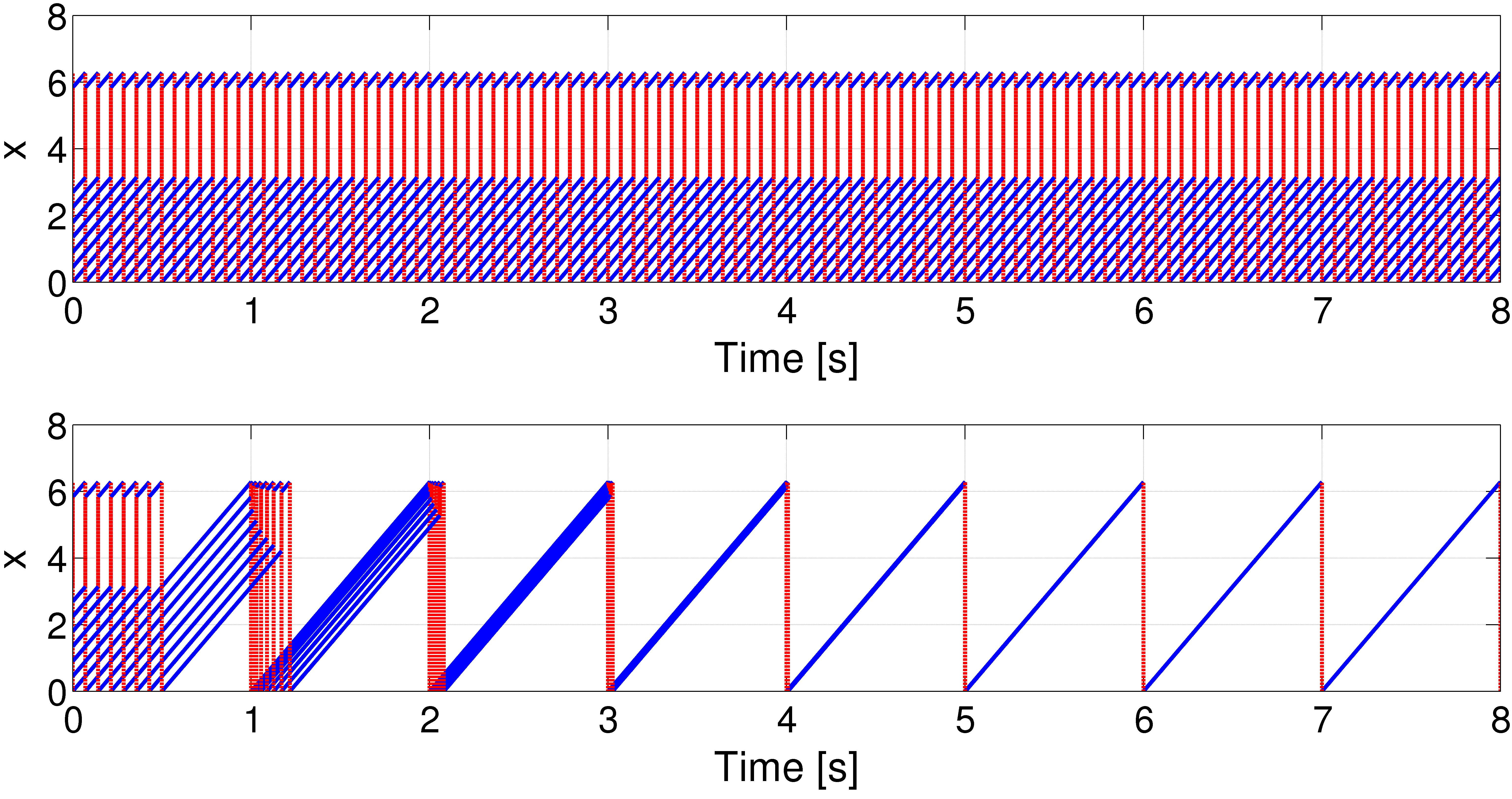}\end{center}
\caption{Simulation results for the \textbf{unidirectional} ring of figure \ref{net.dist}, initial condition $x(0,0)\in\mathcal{U}_1^*$ and there is a refractory period of length $r=\pi$ in node 1. Red lines denote jump instants and blue lines denote phase values. On the top plot $l=0.857<l^*$ hence the network cannot synchronize. On the bottom plot $l=0.86$; since in this case $l>l^*$ the network synchronizes.}% from the worst case.}
\label{res.unidir18}
\end{figure}

\begin{figure}
\begin{center}\includegraphics[
width=0.48\textwidth,keepaspectratio]{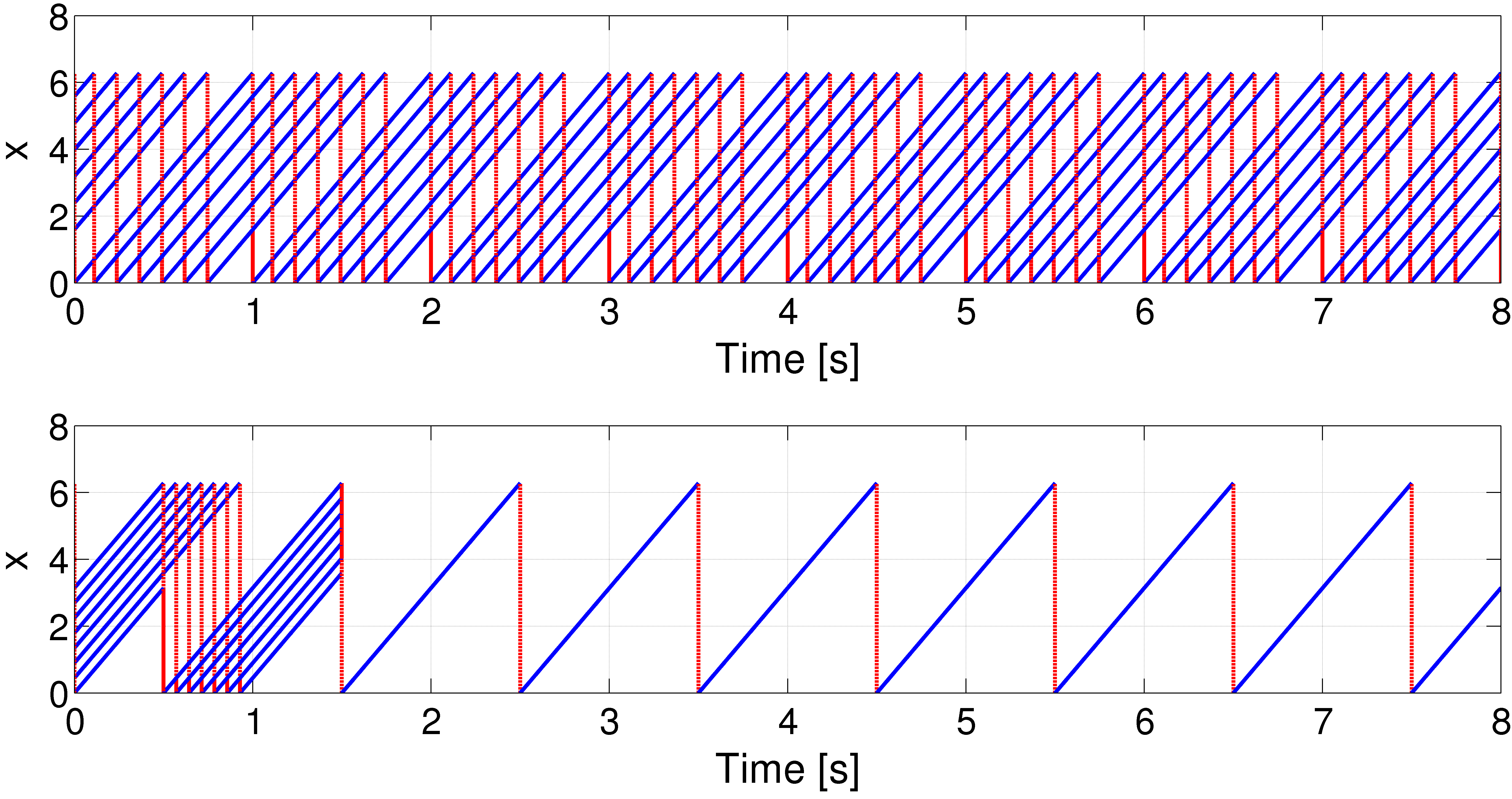}\end{center}
\caption{Simulation results for the \textbf{unidirectional} ring of figure \ref{net.dist} and initial condition $x(0,0)\in\mathcal{U}_2$. Red lines denote jump instants and blue lines denote phase values. On the top plot $l=1$ and there is no refractory period in any node; as was predicted the network cannot synchronize. On the bottom plot $l=1$ and there is a refractory period of length $r=\pi$ in node 1. The network recovers the synchronization properties and synchronizes.}% The network asymptotically synchronizes from the worst case, and thus globally from any initial condition.}
\label{res.unidir81}
\end{figure}

Figure \ref{res.unidir18} shows the results for the unidirectional graph with initial condition $x(0,0)\in\mathcal{U}_1^*$ when there is a refractory period of length $r=\pi$ in the PRC of oscillator 1. Solving the condition in Theorem \ref{thm.2} gives a critical coupling strength of $l^*=0.8571$. It can be seen in the top plot that when $l=0.857<l^*$ the network cannot synchronize and the oscillators distribute in the interval $[0,2\pi]$. Note that $\mathcal{U}_1^*$ is a TDMA-like equilibrium for the system when $l<l^*$.  Increasing the coupling strength such that $l=0.86>l^*$ asymptotically synchronizes the network, as shown in the bottom plot. Figure \ref{res.unidir81} shows the results for the unidirectional graph when the initial condition $x(0,0)\in\mathcal{U}_2$ and there no oscillator is affected by a refractory period. In this case, the network cannot synchronize even when the coupling strength is $l=1$ (the maximum possible value), which is shown in the top plot of Figure \ref{res.unidir81}. The bottom plot shows the results when a refractory period of length $r=\pi$ is introduced in the PRC of oscillator $1$. The network recovers the synchronization properties and synchronizes. It should be noted that since $l=1$, an absorption phenomenon occurs yielding synchronization in finite time.


\begin{figure}
\begin{center}\includegraphics[
width=0.4\textwidth,keepaspectratio]{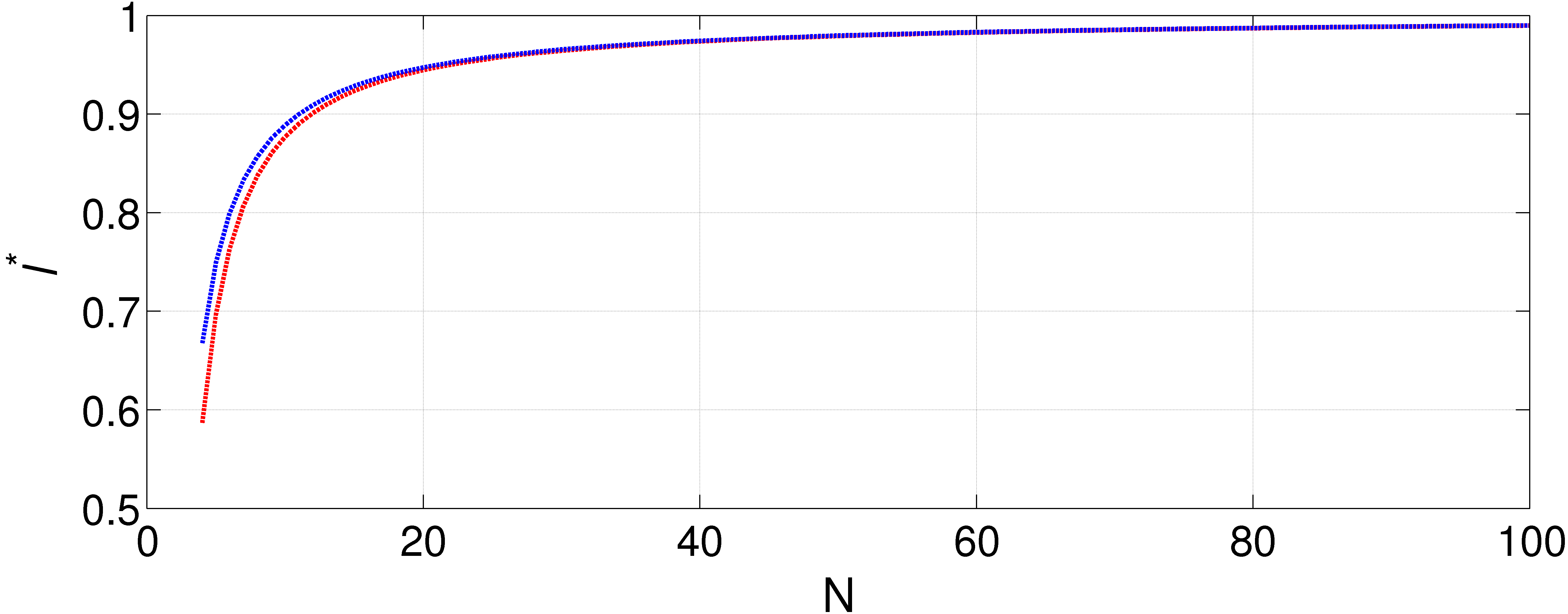}\end{center}
\caption{Critical coupling strength $l^*$ as a function of $N$ for the unidirectional (blue) and bidirectional (red) cases.}
\label{fig.coup}
\end{figure}

Figure \ref{fig.coup} shows the critical strength $l^*$ as a function of the number of oscillators $N$ for both the unidirectional (blue curve) and bidirectional (red curve) cases. It can be seen (also deduced from the condition in the theorems) that $l^*$ is always larger for unidirectional graphs and that, as the number of oscillators grows, the coupling strength goes to the maximal value $1$. In fact, for $N=250$ we have $l^*=  0.99598$ for the unidirectional case, and $l^*=0.99597$ for the bidirectional case.

\section{Conclusions}
\label{sec.5}
Global synchronization of PCOs on cycle graphs is analyzed from a hybrid systems perspective. Using a well-posed hybrid model of a network of PCOs it is possible to formulate synchronization as a set stabilization problem and solve it accordingly. It is proven that global synchronization can be achieved in a bidirectional cycle graph if the coupling strength is above the critical value, while global synchronization in a unidirectional cycle graph cannot be achieved unless a refractory period is introduced in the phase response curve of one of the oscillators. The critical value of the coupling strength was found to increase monotonically with the number of oscillators in the cycle $N$, approaching the maximal value 1 as $N$ goes to infinity. To confirm our theoretical findings, we presented numerical experiments conducted in a general hybrid systems simulator. 

Future work in this line includes extending the proposed approach to the analysis of PCO networks interacting on general graphs, and the analysis of PCO networks coupled through general non-optimal PRCs.

\section*{Appendix}
\appendix
\section{Proof of Proposition \ref{prop.1}}
To prove Proposition \ref{prop.1} we need the following result.
\begin{lem}[Theorem S3 \emph{in \cite{goe:09}}]
\label{lem.1}
Suppose $\tilde{\mathcal{H}}$ is well posed and, for every $\xi\in\mathcal{C}\cup\mathcal{D}$, there exists a nontrivial solution to $\tilde{\mathcal{H}}$ starting from $\xi$. Let $x$ be a maximal solution to $\tilde{\mathcal{H}}$. Then exactly one of the
following three cases holds:
\begin{enumerate}
 \item[(a)] $x$ is complete
 \item[(b)] $x$ blows up in finite hybrid time
 \item[(c)] $x$ eventually jumps out of $\mathcal{C}\cup\mathcal{D}$.
\end{enumerate}
\end{lem}
The proof of Proposition \ref{prop.1} follows.
\begin{pf}
 To analyze existence, note that for every $\xi\in\mathcal{C}\backslash \mathcal{D}$ there exists $\sigma>0$ and an absolutely continuous function $z : [0, \sigma] \to \mathbb{R}^n$ such that $z(0) = \xi$, $\dot{z}(t)= F(z(t))$ for almost all $t\in[0, \sigma]$ and $z(t)\in \mathcal{C}$ for all $t\in (0, \sigma]$. Note also that $G(\mathcal{D})\subset \mathcal{C}\cup \mathcal{D}$. Then there exists a nontrivial solution from every initial condition in $[0,2\pi]^{N}$.

Since $G(\mathcal{D})\subset \mathcal{C}\cup \mathcal{D}$, condition (c) of Lemma \ref{lem.1} is not satisfied. Now it is convenient to point out that since $F(x)=[w,\ldots, w]^T$ is constant, it is globally Lipschitz, and there are no finite escape times. So, no maximal solution can satisfy condition (b) and therefore, all maximal solutions satisfy condition (a) of Lemma \ref{lem.1}, i.e., are complete.

Note that $\cap \mathcal{D}_k\neq \emptyset$, i.e., a point $\tilde{\phi}(t,j)\in[0,2\pi]^{N}$ might belong to more than one $\mathcal{D}_k$. By construction $G(\tilde{\phi})$ is such that if $\tilde{\phi}(t,j)$ belongs to exactly $m$ sets from the collection $\mathcal{D}_k$, with $m\leq N$, then there will be at least $m$ consecutive jumps with no flow in between; moreover, after $m$ jumps it is possible that $\tilde{\phi}(t,j+m)$ belongs to others $\mathcal{D}_k$ due to the coupling effect, and more jumps are required. In any case, there will be at most $N$ consecutive jumps with no flow in between since Assumption \ref{ass.3} gives $Q(0)=0$.  

It follows that the amount of ordinary time between jumps is upper bounded by the natural period of the network, $\tfrac{2\pi}{w}$. To see this, suppose that an oscillator has just fired and every oscillator has phase equal to 0. The next firing will occur after an amount of time equal to $\tfrac{2\pi}{w}$.  If an oscillator has phase larger than 0, it will fire before an amount of time of $\tfrac{2\pi}{w}$ has elapsed unless it receives a pulse in the delay part of the PRC; however, this contradicts the absence of pulses in the network. Then, the upper bound is $\tfrac{2\pi}{w}$. \qed
%
%Finally, it is clear that the minimum oscillating period for a given oscillator $i$ is given by $\tfrac{2\pi-\delta^+ x_i}{w_i}$ where $\delta^+ x_i$ is the maximum phase advance due to pulses. Similarly, the maximum oscillating period is given by $\tfrac{2\pi+\delta^- x_i}{w_i}$.
\end{pf}

\section{Proof of Lemma \ref{lem.v}}
\begin{pf}
Define $\bar{i}:=\min\{i^*,i_*\}$. To prove statement $(a)$  note that $|x_{i}-x_{i+1}|\leq\pi\Rightarrow V_i=|x_{i}-x_{i+1}|$. If we add the $V_i$s, the minimum is reached when the segments are disjoint, i.e., phases are ordered either clock-wise, or counter clock-wise, and in this case  $\sum_{i\in\mathcal{V}\setminus\{\bar{i}\}}V_i= |x_{i^*}-x_{i_*}|$. Now we proceed to prove by contraposition. Suppose $|x_{i}-x_{i+1}|\leq\pi\,\forall i\in\mathcal{V}\setminus\{i^*,i_*\}$ and $ |x_{i^*}-x_{i_*}|\geq\pi$. Then, the minimum length of the cycle is equal to $\sum_{i\in\mathcal{V}\setminus\{\bar{i}\}}V_i+\min\left(|x_{i^*}-x_{i_*}|,2\pi-|x_{i^*}-x_{i_*}|\right)$, which corresponds to the component measuring the length between $x_{i^*}$ and $x_{i_*}$ (note that for the length to be minimal, ${i^*}$ and ${i_*}$ must be neighbors). Then we have $\mathbf{1}^TV\geq\sum_{i\in\mathcal{V}\setminus\{\bar{i}\}}V_i+2\pi- |x_{i^*}-x_{i_*}|=2\pi$. Hence, $(a)$ holds.

Regarding $(b)$, note that for the length to be larger than $2\pi$ segments cannot be disjoint since from the previous paragraph we know that disjoint segments can add up to $2\pi$, the length of the domain. Then, there must be the case that at least 2 segments, described by the components of $V$, intersect. Considering $x_i=2\pi$, the conditions  $x_{i+2}\in[0,x_{i+1})$ and $x_{i+1}\leq\pi$, or $x_{i+2}\in(x_{i+1},2\pi]$ and $x_{i+1}\geq\pi$, or  $|x_{i+2}-x_{i+1}|>\pi$ ensure that at least 2 segments affected by $i$ intersect. Moreover, these conditions imply that the length of the cycle, $\mathbf{1}^TV$, will decrease after $i$ jumps.

In the same line, when $\mathbf{1}^TV=2 \pi$ and $x\notin\mathcal{U}$ segments are not disjoint and then there exists $i$ such that when $x_i=2\pi$, we have $x_{i+2}\in[0,x_{i+1})$ and $x_{i+1}\leq\pi$, or $x_{i+2}\in(x_{i+1},2\pi]$ and $x_{i+1}\geq\pi$, or  $|x_{i+2}-x_{i+1}|>\pi$ holds, implying that the length of the cycle, $\mathbf{1}^TV$, will decrease after $i$ jumps. Hence, $(c)$ holds.

Statement $(d)$ follows by noting that the phase ordering implies that $i^*$ and $i_*$ are neighbors. Moreover,  from $\mathbf{1}^TV=2 \pi$  we have $\sum_{i\in\mathcal{V}\setminus\{\bar{i}\}}V_i= |x_{i^*}-x_{i_*}|$ or $\sum_{i\in\mathcal{V}\setminus\{\bar{i}\}}V_i= 2\pi-|x_{i^*}-x_{i_*}|$ depending on whether $|x_{i^*}-x_{i_*}|\geq\pi$ or $|x_{i^*}-x_{i_*}|<\pi$ holds. Suppose the latter is true and then $|x_i-x_{i+1}|<\pi$ holds for every oscillator $i\in\mathcal{V}$; furthermore, the phase ordering implies that segments are disjoint and then $\sum_{i\in\mathcal{V}\setminus\{\bar{i}\}}V_i= |x_{i^*}-x_{i_*}|$, which contradicts $\mathbf{1}^TV=2 \pi$. Hence $|x_{i^*}-x_{i_*}|\geq\pi$ must hold. Now if $|x_{i^*}-x_{i_*}|\geq\pi$ holds, either $|x_i-x_{i+1}|\leq\pi$ holds for every oscillator $i\in\mathcal{V}\setminus\{\bar{i}\}$ or $|x_i-x_{i+1}|>\pi$ holds for only one oscillator $i$ (due to the phase ordering). Suppose the latter is true (note that for this to be feasible $|x_{i^*}-x_{i_*}|>\pi$ must hold),  then we have $\sum_{i\in\mathcal{V}\setminus\{\bar{i}\}}V_i= 2\pi+|x_{i^*}-x_{i_*}|-2|x_i-x_{i+1}|<|x_{i^*}-x_{i_*}|$, which again contradicts $\mathbf{1}^TV=2 \pi$. Hence, $|x_i-x_{i+1}|\leq\pi$ must hold for every oscillator $i\in\mathcal{V}\setminus\{i^*,i_*\}$ and the Lemma is proven. \qed
\end{pf}
%\section{Proof of Lemma}
%\section{Proof of Lemma}

\bibliographystyle{plain}
\bibliography{biblio}

\end{document}